# Ferroelectric AlScN thin films with enhanced polarization and low leakage enabled by high-power impulse magnetron sputtering

Federica Messi[1], Jyotish Patidar[1], Nathan Rodkey[1], Christoph W. Dräyer[2], Morgan Trassin[2], Sebastian Siol[1,*]

[1]Laboratory for Surface Science and Coating Technologies,
Empa – Swiss Federal Laboratories for Materials Science and Technology, Switzerland

[2]Department of Materials, ETH Zürich, Switzerland

E-mail: sebastian.siol@empa.ch

## Abstract

The demand for efficient data processing motivates a shift toward in-memory computing architectures. Ferroelectric materials, particularly AlScN, show great promise for next-generation memory devices. However, their widespread application is limited due challenges such as high coercive fields, leakage currents and limited stability. Our work introduces a novel synthesis approach for ferroelectric AlScN thin films using high-power impulse magnetron sputtering (HiPIMS). Through a combinatorial study, we investigate the effect of scandium content and substrate bias on the ferroelectric properties of AlScN films deposited using metal-ion synchronized (MIS) HiPIMS. Leveraging the high ionization rates of HiPIMS and optimally timed substrate bias potentials, we enhance the adatom mobility at low temperatures. Our films exhibit a high degree of texture and crystallinity as well as low roughness at temperatures as low as 250°C. Most importantly, the films exhibit coercive fields comparable to state-of-the-art values (5 MV/cm) with significantly enhanced remanent polarization (158-172.0 $\mu C/cm^2$). Notably, the remanent polarization remains stable across varying scandium concentrations. We further evaluate cycling stability and leakage current to assess suitability for memory applications. This study demonstrates HiPIMS as a scalable and CMOS compatible technique for synthesizing high-quality ferroelectric AlScN films, paving the way for their application in non-volatile memory applications.

# Introduction

The development of artificial intelligence (AI) and machine learning have sparked growing demand for rapid and efficient data processing, motivating innovation in computing architectures. Central to this evolution is the advancement of memory technologies, particularly in-memory computing devices, which offer solutions to overcome bottlenecks associated with traditional von Neumann computer architecture. Among these, non-volatile memory technologies such as Ferroelectric Random Access Memories (FeRAMs) and Ferroelectric Field-Effect Transistors (FeFETs) hold significant promise due to their low power consumption, durability and cycling speeds. [1]–[3] This functionality is enabled by a ferroelectric materials' inherent ability to maintain its polarization after the removal of an electric field. [4] Perovskite ferroelectrics and the recently established ferroelectric hafnium oxide system have been widely studied for this purpose. [5]–[7] However, the limited compatibility of perovskites oxides with complementary metal-oxide semiconductor (CMOS) technology due to their demanding, high temperature deposition processes pose significant challenges for device integration. [3], [8], [9] While, significant improvements have been made in hafnia-based technologies, its limited remanent polarization as well as pronounced wake-up and fatigue behavior still limit their wide-scale applicability. [6], [7], [10], [11] Recently, aluminum scandium nitride ($Al_{1-x}Sc_xN$) has emerged as a promising candidate for ferroelectric applications. [12]–[18] Discovered to be ferroelectric in 2019 by Fichtner *et al.*, $Al_{1-x}Sc_xN$ showed remnant polarization exceeding 100 µC/cm² and a decreasing coercive field with increasing Sc content. [19] Wolff *et al.* further confirmed its ferroelectric nature using STEM imaging, demonstrating clear atomic-scale polarization inversion between Al-polar and N-polar states. [20] These findings have sparked a wave of research aimed at optimizing the material's properties, including alloying with scandium, film thickness, and leakage current reduction. [19], [21]–[25] Thickness-dependent effects present significant challenges as polarization diminishes in thinner films, with hysteresis loops disappearing entirely below 20 nm. [24], [26] This thickness limit was later overcome by Schönweger *et al.* who used molecular beam epitaxy (MBE) to deposit sub-5 nm films on n-GaN/sapphire and Mo (011) substrates, achieving ferroelectric switching at 1 V. [22], [25] Moreover, improvements in deposition methods have significantly reduced leakage currents, especially in epitaxially grown thin films. [21], [24], [27], [28]

The deposition technique plays a deciding role in determining the crystal structure, texture, and overall performance of ferroelectric $Al_{1-x}Sc_xN$ films. [13], [22], [23], [29]–[32] As the performance of ferroelectric devices is directly coupled with texture and microstructure, process development is an integral part in the development of better ferroelectrics. Parameters such as remnant polarization depend on the c-axis alignment of grains, while compact microstructures and controlled stress states can improve switching performance and reduce leakage. [19], [29]–[31]

Recent studies reveal a clear trend of reduced coercive field ($E_c$), but unfortunately also remanent polarization ($P_r$), with increased Sc concentration, regardless of the deposition process. [14],[18],[25],[26],[27] Different deposition methods such as RF sputtering, MOCVD, and MBE have demonstrated varying effects on ferroelectric properties. Yasuoka *et al.* found that RF sputtering results in higher coercive fields, while MOCVD and MBE yield higher remanent polarization, as shown by Wolff *et al.*, and Wang *et al.*. [15], [33], [34] Despite their advantages in producing high-quality thin films, MOCVD and MBE face challenges. Both techniques typically require high deposition temperatures limiting CMOS compatibility, while MBE in addition is limited in scalability. [2], [3], [35], [36] These challenges underscore the need for deposition techniques that combine scalability, CMOS compatibility, and precise control over film properties. Since a ferroelectrics' switching behavior depends on multiple parameters, high-throughput approaches, such as combinatorial screening, can significantly accelerate the development of new ferroelectric materials. [37], [38]

In this paper, we utilize a combinatorial screening workflow combined with an innovative deposition method for the accelerated development of high-quality ferroelectric $Al_{1-x}Sc_xN$ thin films. We demonstrate that High Power Impulse Magnetron Sputtering (HiPIMS) is a highly effective solution for depositing ferroelectric $Al_{1-x}Sc_xN$ films with excellent performance. By applying power in short pulses, HiPIMS processes exhibit particularly high plasma densities, enhancing ionization and kinetic energies. The resulting increase in adatom mobility results in compact, highly textured films at reduced deposition temperatures. [39], [40] The ionized flux fraction can be further accelerated by applying a negative substrate potential, adding an additional degree of freedom in the process parameter space. [41] In recent years, the technique has been further advanced by introducing the ability to tailor the kinetic energy of specific ionic species. [42] In Metal-Ion Synchronized-HiPIMS (MIS-HiPIMS) the negative substrate bias potential is synchronized with the arrival of metal ions at the substrate position. Using this approach, one can selectively accelerate metal ions onto the substrate, reducing surface roughness, tailoring stress states, and minimizing process gas incorporation. [39], [43] This approach ensures CMOS compatibility while offering scalability, cost-effectiveness, and exceptional film quality. Building on our success in piezoelectric thin films synthesis with MIS-HiPIMS, we extend this technique to the deposition of ferroelectric $Al_{1-x}Sc_xN$. [39], [43] Here, we use a high-throughput screening workflow to elucidate the influence of composition, thickness and ion kinetic energy on the structural and functional properties of ferroelectric $Al_{1-x}Sc_xN$ thin films deposited using HiPIMS. The results of this work show that our scalable, low-temperature process yields $Al_{1-x}Sc_xN$ films with remarkable quality and performance. In particular, a strong improvement in the remnant polarization is achieved, which is in line with the observed increase in textured volume-fraction of the films. In addition, the leakage current densities are among the lowest reported to date. Comparable results have previously only been achieved via epitaxial growth using MBE or MOCVD. This highlights how this new deposition approach could pave the way for scalable back-end-of-line (BEOL) deposition of high-quality ferroelectric thin films. [15], [34]

## Methods

The deposition of the libraries is performed using HiPIMS in a commercial, custom-built chamber (AJA International, ATC-1800). The films are deposited on p-type (001), 2x2 in$^2$ silicon wafers. A 9 x 5 grid defines 45 samples on each library for the combinatorial materials libraries. For the depositions, two sputter guns are used at a relative angle of 90° resulting in both thickness and composition gradients on the library. The guns are equipped with an Al target (2" dia., HMW Hauner, purity: 99.9999 at.%) and a Sc target (2" dia, Plasmaterials, purity: 99.9 at.%, oxygen content <1000 ppm), respectively. Hysteresis studies are reported in **SI 1**. The unbalanced magnetrons are oriented in an open-field magnetic configuration. The substrate is placed at a working distance of 12 cm and not rotated during the deposition. The deposition of all the layers is carried out at 250°C, the gas flow is set to 20 sccm for Ar and 12 sccm for $N_2$ and the working pressure is kept at 5 µbar. Before the deposition, the Si substrates are etched by RF Ar plasma at 30 W for 8 min to remove the native oxide layer. Secondly, a 10 nm titanium layer is deposited to ensure good adhesion, thirdly a 100 nm Pt layer is deposited as a back contact and finally our $Al_{1-x}Sc_xN$ films. Four libraries are synthetized with different substrate bias potentials: grounded (i.e. 0 V), -5 V, -10 V and -20 V. The square platinum (Pt) top contacts have edge lengths of 200 µm, 400 µm, and 600 µm. The power and current densities for DCMS and HiPIMS for each sputter gun, during the deposition of combinatorial libraries are summarized in **Table 1**, while the set-up of the deposition chamber is illustrated in **Figure S2**.

*Table 1.* Power- and current densities for Ti, Pt, Al, and Sc sputter guns operated in DCMS and HiPIMS mode during the deposition of the combinatorial libraries.

|  | Power density (W/cm$^2$) | Current density (mA/cm$^2$) |
|---|---|---|
| **Ti - DCMS** | 5 | 16 |
| **Pt - DCMS** | 2 | 8 |
| **Al - HiPIMS** | 2.7 | 350-450 |
| **Sc - HiPIMS** | 2 | 250-350 |

The synchronization of the HiPIMS pulses on both the targets and the substrate is shown in **Figure 1**. Power is applied to both targets in 10 µs pulses at a frequency of 5 kHz, The substrate bias, when applied, is synchronized to the arrival of metal ions at the substrate. This evaluation is done with gated mass spectroscopy measurements (Hiden EQP 300) to study the time of flight of different ion species present in the plasma. More details regarding the measurement are provided in our previous works. [39]

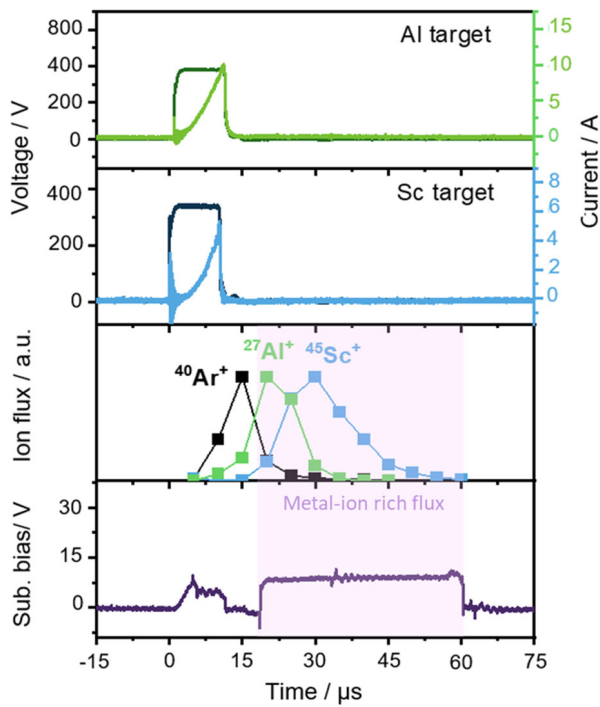

**Figure 1.** *Synchronization of substrate bias pulse. First, the pulses on Al and Sc targets are synchronized to start at the same time. Then, a negative substrate bias potential is synchronized to the metal-rich part of the HiPIMS flux as determined by gated mass spectrometry.*

X-ray diffraction (XRD) analysis of the films was conducted using a Bruker D8 diffractometer in Bragg-Brentano geometry with Cu-Kα radiation. θ-2θ scans were used to assess the crystal orientation of the samples, covering a 2θ range from 30° to 45° with a step size of 0.01°. Additionally, rocking curve measurements were carried out for the (002) peak over a ω range of 12° to 24° to evaluate the c-axis texture of the films. The thickness of the samples was measured using a Dektak profilometer. The surface composition of the films was determined using X-ray photoelectron spectroscopy (XPS) using a PHI-Quantera with monochromated Al kα radiation. The cation ratio is calculated based on the area of the Al 2s spectra is compared to the area of the Sc 2s spectra after Shirley background subtraction considering the respective relative sensitivity factors.

The ferroelectric properties of all library samples were measured using a thin film analyzer (aixACCT, TF2000). The ferroelectric characterization of the samples starts with a dynamic hysteresis measurement where the samples are subjected to a voltage in the range of 120-200 V at a frequency of 1 kHz. In all reported results, the coercive field is defined as the average of the positive and negative fields at which the current density reaches its maximum. Similarly, the remanent polarization is calculated as the average of the positive and negative polarization values at 0 V/cm. The $P_r$ and $E_c$ values are also corrected with respect to the leakage current through a PUND (Positive-Up Negative-Down) measurement where the contribution from leakage current can be isolated from the switching current. Secondly, a fatigue measurement is carried out. Here the samples are cycled at 10 kHz at the switching and at the reading potential to observe how the polarization evolves during cycling and estimate their cycling endurance.

## Results and discussion

### Structural characterization of the films

In this work we investigate the potential benefits of High-Power Impulse Magnetron Sputtering (HiPIMS) for the synthesis of high-quality ferroelectric $Al_{1-x}Sc_xN$ thin films. To this end, combinatorial libraries were synthesized with varying scandium content (8-28% **Figure S3**) and film thickness (169-340 nm). First, X-ray diffraction analysis is performed on $Al_{1-x}Sc_xN$ materials libraries synthesized for different biasing conditions (see **Figure S4-S5**). We find that all synthesized samples crystallize in wurtzite structure with a pronounced (002) out-of-plane texture. The results of the structural analysis are summarized in **Figure 2**. **Figure 2a** shows XRD scans of representative films with an approximate Sc concentration of x = 0.22 and a thickness of approximately 260 nm. The films have a high degree of crystallinity and pronounced c-axis texture. A minor shift of the $Al_{1-x}Sc_xN$ (002) peak towards lower 2θ values is observed for higher substrate biases, which can be explained by film densification upon metal-ion bombardment. Given the critical importance of texture in determining the overall crystal quality and, by extension, the functional properties of $Al_{1-x}Sc_xN$ films, a quantitative assessment of the texture is essential. [9], [26], [44]–[50] To this end, the rocking curve for the $Al_{1-x}Sc_xN$ (002) peak was measured on all samples. **Figure 2b and 2c** reveal that samples biased at -10 V exhibit the lowest Full Width at Half Maximum (FWHM) values, indicating superior out-of-plane texture and lower mosaicity. The FWHM trends across different bias conditions show that -10 V yields the best texture, followed by -5 V with intermediate FWHM values, while both grounded (0 V) and -20 V biased libraries display similar, higher FWHM values. These results suggest that moderate substrate bias (-10 V) is ideal to enhances the adatom mobility of the film forming species without introducing excess compressive stress or defects. These results were further confirmed on $Al_{0.78}Sc_{0.22}N$ samples synthesized on smaller wafers with substrate rotation. Stress analysis on these samples clearly demonstrates the change in residual stress for different biasing conditions. While the samples deposited without any additional ion acceleration exhibit tensile stress on the order of 100 MPa, samples deposited with -10 V and -20 V showed approximate stress values of 0 MPa and -100 MPa, respectively (see **Figure S6**). AFM analysis revealed low surface roughness (RMS 1.27 to 2.58 nm over 4 µm²) for all films, significantly lower than typical DCMS values and comparable to HiPIMS and MBE methods, with the lowest roughness observed in films deposited with a synchronized substrate bias of -10 V. (see **Figure S7**) This enhancement in surface quality is indicative of improved texture and reduced Full Width at Half Maximum (FWHM) associated with -10 V substrate bias deposition condition.

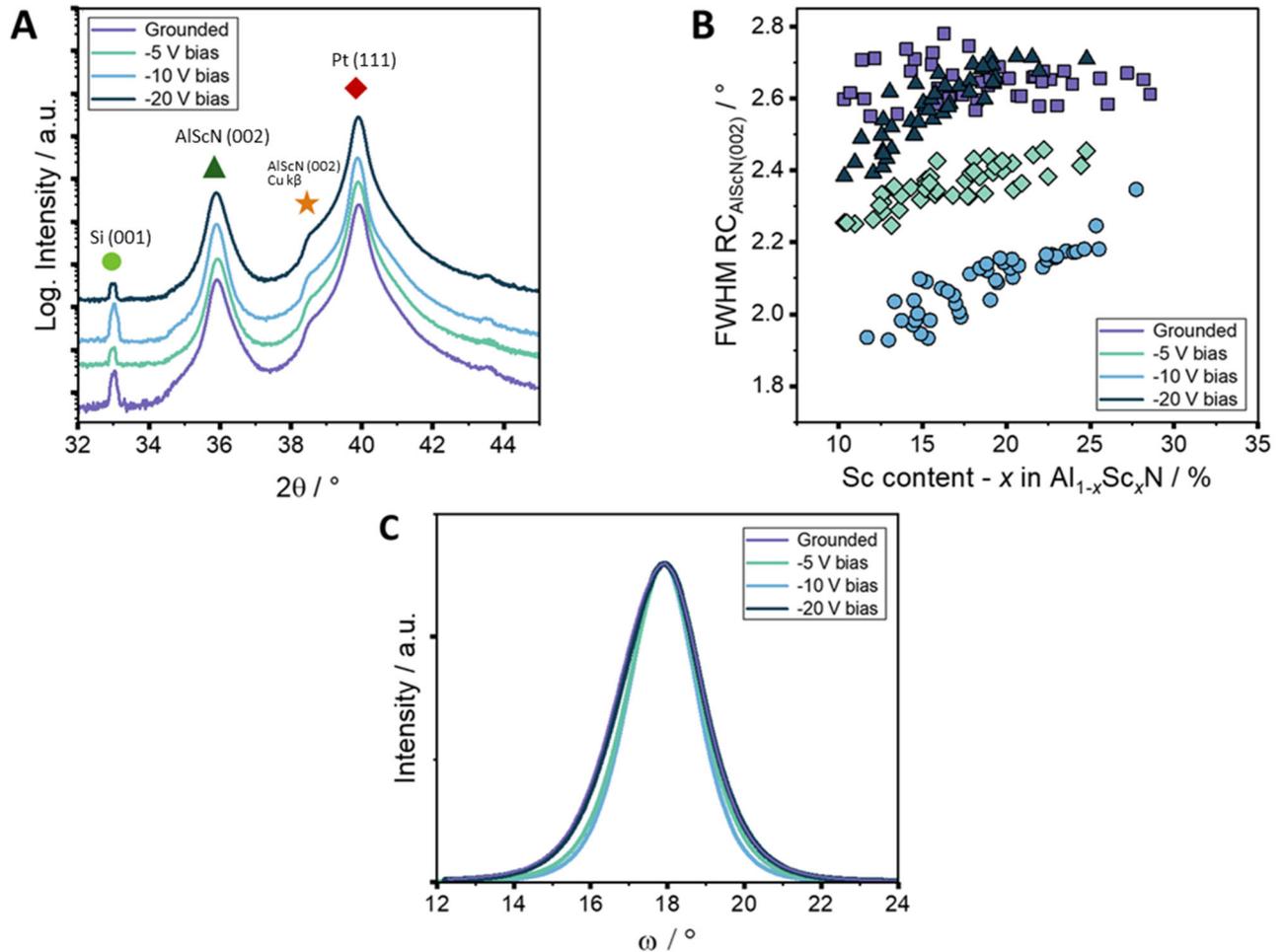

***Figure 2.*** XRD results for $Al_{1-x}Sc_xN$ materials libraries deposited using different bias-potentials (grounded; with -5 V bias; with -10 V bias; with -20 V bias): **a.** θ-2θ measurement of 4 samples of the different libraries with 22% cat.% Sc and 260 nm thickness; **b.** FWHM of the rocking curve of the $Al_{1-x}Sc_xN$ (002) peak as a function of scandium content; **c.** normalized rocking curve of the $Al_{1-x}Sc_xN$ (002) peak of 4 samples with 22 cat.% Sc and 260 nm thickness. These results show improved texture for the library deposited with -10 V substrate bias.

## Ferroelectric characterization

Following the structural analysis on the $Al_{1-x}Sc_xN$ materials libraries, two libraries were selected for further investigation of the ferroelectric properties: the -10 V substrate bias library and the grounded library. The – 10 V library was chosen due to its superior texture and c-axis orientation, which are expected to enhance ferroelectric remanent polarization. The grounded library serves as a reference and it is anticipated to exhibit the highest tensile stress conditions, potentially advantageous for lowering the coercive field. Ferroelectric switching was investigated across all library samples. The range of thicknesses exhibiting ferroelectric switching was not sufficiently broad to comprehensively assess thickness-dependent effects on ferroelectric properties.

Therefore, the analysis focused on a defined thickness range to enable a more meaningful investigation of ferroelectric behavior.

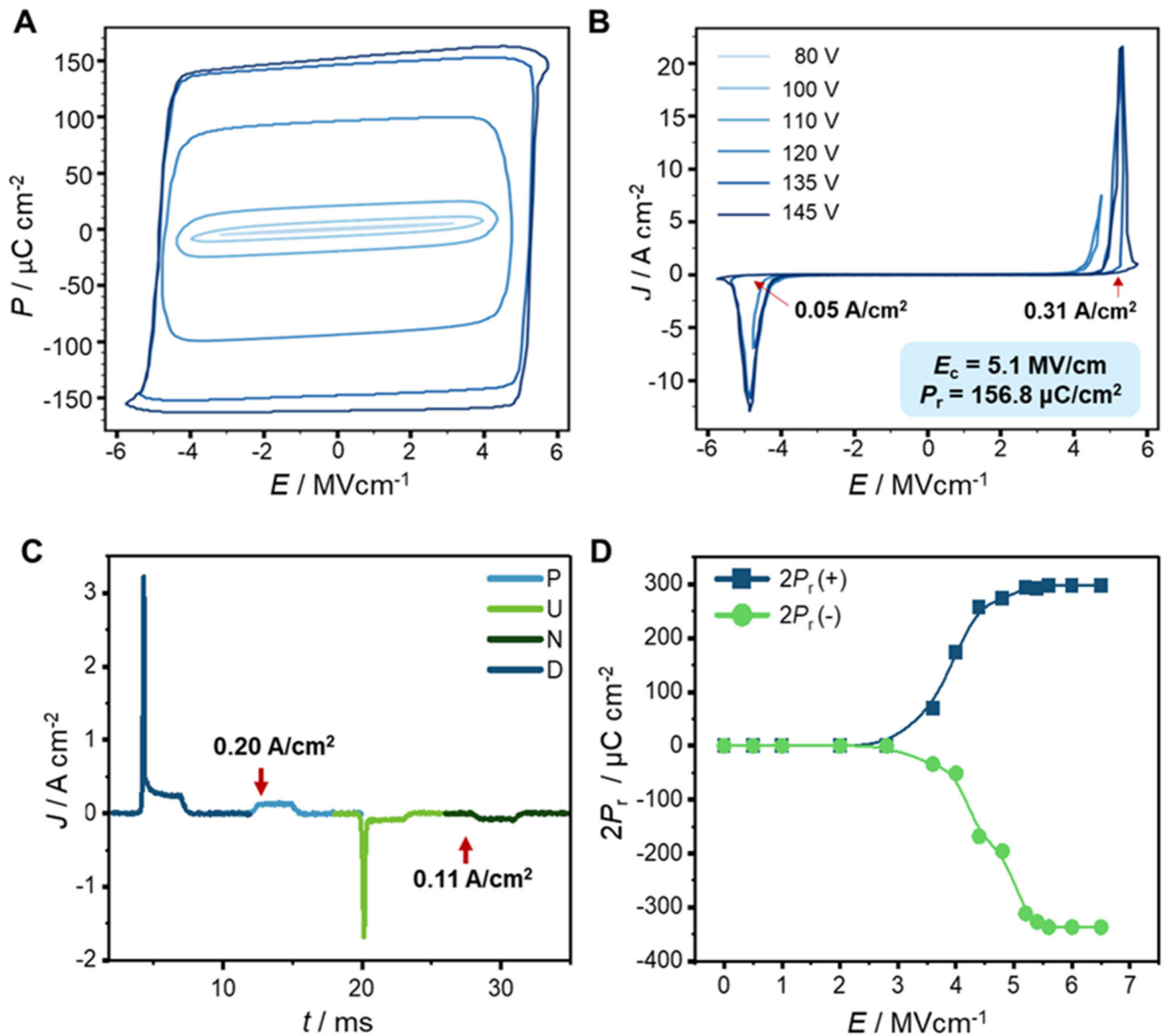

*Figure 3. a.* Nested polarization-electric field (P-E) loops for $Al_{1-x}Sc_xN$ film with 18.5% Sc cat.% Sc and 250 nm thickness, measured using a 1 kHz triangular excitation waveform. *b.* Corresponding current loops demonstrating ferroelectric switching behavior and switching current, with values of the leakage current at the coercive field. (c, d) PUND measurement of $Al_{1-x}Sc_xN$ film with 18.5% cat.% Sc and 250 nm thickness: *c.* current density vs time with reported intensity of the leakage current during the U and D phases; *d.* change in polarization as the potential applied increases to evaluate the true remanent polarization without leakage contribution.

Initial analysis of polarization-electric field (P-E) loops revealed progressive opening with increasing applied voltages until saturation was achieved, as illustrated in **Figure 3.a and 3.b**. Calculated $P_r$, $E_c$ and leakage current values for all switched samples are provided in the supporting information (**Table S1**). When evaluating remanent polarization, it is crucial to address and assess the impact of leakage current, as it can significantly influence the apparent ferroelectric properties of the material. Two effective methods were employed to estimate leakage current: Firstly, analyzing the switching current density at the coercive field (**Figure 3.b**) and secondly, Positive-Up Negative-Down (PUND) measurements (**Figure 3.c and 3.d**). We used both methods on a few selected samples, and the resulting data demonstrated good agreement, as shown in **Figure 3.b and 3.c**. As illustrated in **Figure 3.d**, the true remnant polarization was found to be very similar to the apparent remanent polarization. This observation allowed for the assumption that, for the remaining samples, the apparent remanent polarization could be considered equivalent to the real remanent polarization, provided that the leakage current densities at the coercive field were small.

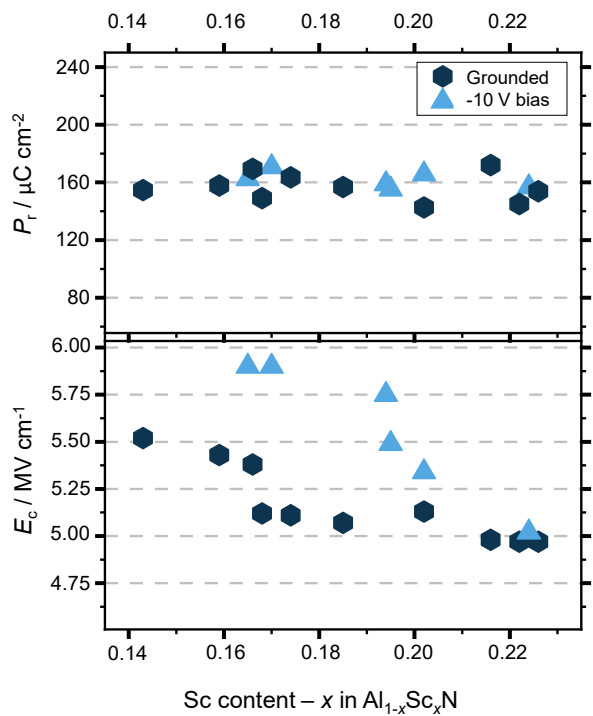

***Figure 4.*** *Variation of the coercive field and the remanent polarization as a function of Sc content, FWHM and substrate bias for the library samples.*

**Figure 4** presents a comprehensive analysis of remanent polarization and coercive field for all switched samples as a function of scandium concentration. The data reveals that texture variations between grounded and biased samples do not significantly impact remanent polarization. (see **Figure 2**) However, stress differences between the libraries indicate lower coercive fields and leakage currents for the more tensile, grounded films. Notably, compared to other $Al_{1-x}Sc_xN$ films reported in literature the rocking curve values are small for both biasing conditions. Strikingly, the films' texture, and consequently the remanent polarization, remain high and are independent of scandium concentration. This is in contrast with results obtained using other deposition methods and highlights the ability of HiPIMS to produce thin-films with consistently pronounced texture. [19], [23], [30] The measured coercive fields, ranging within 5-6 MV/cm, align with average values reported in literature. The grounded substrate bias conditions exhibit an overall slightly lower coercive field, which can be explained by the more tensile residual stress values, as also confirmed by the single composition samples analysis. (see **Figure S6**) This is in line with observations by Fichtner *et al.,* who demonstrated the importance of both Sc alloying as well as tensile strain to lower the barrier for spontaneous polarization. [19]

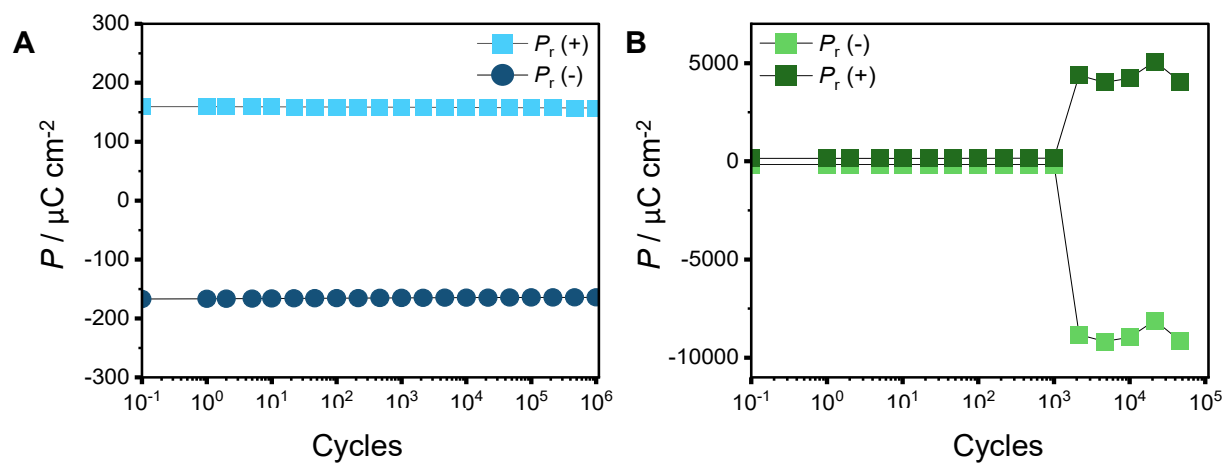

*Figure 5. Cycling stability of the library sample with 18.5% cat.% Sc and a thickness of 250 nm: **a.** read cycling (110 V) showed stability up to $10^6$ cycles (110 V); **b.** write/switching cycling (140 V) showed stability up to $10^3$ cycles.*

Finally, cycling endurance tests were conducted to evaluate read and write stability, crucial metrics for memory applications. This measurement consists in applying a series of voltages, either above or below the coercive field of the sample, to respectively read or switch the polarization state, while monitoring the remanent polarization after multiple cycles. Details regarding the testing protocol are provided in the supporting information. Read-cycles (**Figure 5a**) of a sample with 18.5 cat.% Sc and 250 nm thickness were stable after $10^6$ cycles at a read voltage of approximately 110 V. The test was concluded without any signs of device degradation. However, switching stability tests (**Figure 5b**) revealed limited durability, with samples experiencing sudden shorts after about $10^3$ cycles. Despite the apparent room for improvement these results are among the most promising values reported for $Al_{1-x}Sc_xN$ to date. [28], [51]–[54] The absence of a gradual decrease in polarization over cycles suggests that the failure was caused by defects and charge accumulation, leading to a short in the sample, indicating that improving the quality of the top contacts could potentially enhance stability.

To evaluate the potential of the process to reduce film thickness and consequently the switching potential samples with 10 nm film thickness and 24 cat.% Sc were synthesized with the same overall bottom contact and characterized by piezo force microscopy (PFM). This scanning probe technique allows for the investigation of local switching events and hence, is less sensitive to detrimental leakage current contributions in the assessment of the switchability of the thinner layers. The PFM results showed ferroelectric switching < 10 V as shown in **Figure S8**. While further process improvement is needed to manufacture ultra-thin ferroelectric devices with acceptable yield, the initial results are more than promising and highlight the great potential of ionized physical vapor deposition processes such as HiPIMS for the deposition of ferroelectric thin films.

## Summary and Conclusion

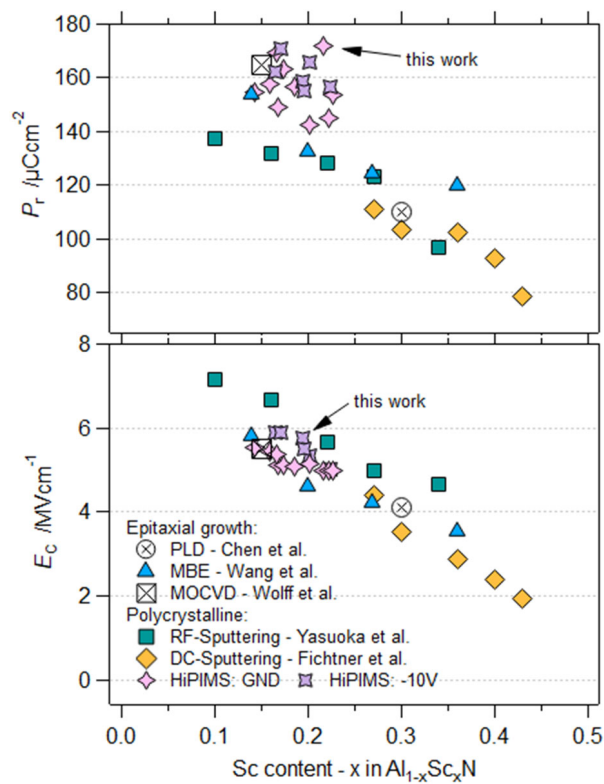

***Figure 6.*** *Comparison of libraries results with state-of-the-art results of $P_r$ and $E_c$ as a function of scandium content. The results reveal that the achieved $E_c$ values are comparable to the best available results, while the $P_r$ values are strikingly high.*

This study demonstrates the efficacy HiPIMS for synthesizing highly crystalline $Al_{1-x}Sc_xN$ thin films for ferroelectric applications. The increased adatom mobility achievable in this process enables enhanced ferroelectric performance at comparatively low synthesis temperatures. **Figure 6** shows a comparison of our HiPIMS-de-posited films with values previously reported in literature. The HiPIMS-deposited Al$_{1-x}$Sc$_x$N films exhibit excellent remanent polarization values independent of scandium concentration and substrate bias. Comparable values have so far only been achieved using heteroepitaxial growth using MBE or MOCVD. [30], [32] This increase in performance compared to conventional sputtering can be explained by an increased in the improvement in c-axis oriented volume in the films. Coercive field results were comparable to state-of-the-art values, with a confirmed trend of decreasing coercive field with increasing scandium content and tensile stress. [19], [23], [30]–[32] Excellent read cycling stability up to $10^6$ cycles was demonstrated, although write cycling stability requires further improvement. Single-composition sample analysis corroborated the predicted stress states and surface roughness observed in the combinatorial libraries, validating our approach. It is noteworthy that these results were achieved using comparably low deposition temperatures of only 250 °C, fully compatible with the back-end of line (BEOL) requirements towards integration of the industry-relevant Silicon platform. Recent progress in the low-temperature growth of highly-textured AlN and AlScN indicate that this value could be further reduced in the future enabling the deposition of ferroelectric thin films on temperature sensitive substrates without heating. [55], [56] This study advances our understanding of the relationship between the specific synthesis environments and ferroelectric properties in Al$_{1-x}$Sc$_x$N thin films, paving the way for the development of high-performance ferroelectric nitrides using scalable and CMOS compatible processes.

## Acknowledgements

F.M and J.P. acknowledge funding by the SNSF (Project No. 200021_196980). M.T. acknowledged the Swiss National Science Foundation under Project No. 200021_188414. F.M. would like to thank Kerstin Thorwarth and Oleksandr Pshyk for proving help with XPS training and assistance with the depositions. Help from Alexander Wieczorek for the image design and data analysis training is gratefully acknowledged.

## Author Contributions

Federica Messi: Conceptualization, Investigation, Formal analysis, Visualization, Writing – original draft. Jyotish Patidar: Conceptualization, Investigation, Formal analysis, Supervision, Writing – review & editing. Nathan Rodkey: Investigation, Writing – review & editing. Christoph W. Dräyer: Investigation, Formal analysis, Writing – review & editing. Morgan Trassin: Conceptualization, Investigation, Formal analysis, Writing – review & editing. Sebastian Siol: Conceptualization, Supervision, Methodology, Formal analysis, Funding acquisition, Writing – review & editing.

# Bibliography

[1] P. Wang, D. Wang, S. Mondal, M. Hu, J. Liu, and Z. Mi, "Dawn of nitride ferroelectric semiconductors: from materials to devices," *Semicond. Sci. Technol.*, vol. 38, no. 4, p. 043002, Mar. 2023, doi: 10.1088/1361-6641/ACB80E.

[2] Y. Zhang, Q. Zhu, B. Tian, and C. Duan, "New-Generation Ferroelectric AlScN Materials," *Nano-Micro Lett.*, vol. 16, no. 1, p. 227, Dec. 2024, doi: 10.1007/s40820-024-01441-1.

[3] T. Mikolajick *et al.*, "Next generation ferroelectric materials for semiconductor process integration and their applications," *J. Appl. Phys.*, vol. 129, no. 10, p. 100901, Mar. 2021, doi: 10.1063/5.0037617.

[4] M. Müller, I. Efe, M. F. Sarott, E. Gradauskaite, and M. Trassin, "Ferroelectric Thin Films for Oxide Electronics," *ACS Appl. Electron. Mater.*, vol. 5, no. 3, pp. 1314–1334, 2023, doi: 10.1021/acsaelm.2c01755.

[5] K. H. Cho *et al.*, "Leakage current mechanism and effect of oxygen vacancy on the leakage current of Bi5Nb3O15 films," *J. Eur. Ceram. Soc.*, vol. 30, no. 2, pp. 513–516, 2010, doi: 10.1016/j.jeurceramsoc.2009.04.029.

[6] Z. Zhao *et al.*, "Engineering Hf0.5Zr0.5O2Ferroelectric/Anti- Ferroelectric Phases with Oxygen Vacancy and Interface Energy Achieving High Remanent Polarization and Dielectric Constants," *IEEE Electron Device Lett.*, vol. 43, no. 4, pp. 553–556, 2022, doi: 10.1109/LED.2022.3149309.

[7] T. S. Böscke, J. Müller, D. Bräuhaus, U. Schröder, and U. Böttger, "Ferroelectricity in hafnium oxide thin films," *Appl. Phys. Lett.*, vol. 99, no. 10, p. 102903, Sep. 2011, doi: 10.1063/1.3634052/121850.

[8] J. F. Scott, "Applications of modern ferroelectrics," *Science (80-. )*, vol. 315, no. 5814, pp. 954–959, Feb. 2007, doi: 10.1126/SCIENCE.1129564/ASSET/8BA9DD7F-A0CD-42F8-A874-D9699E6BEF75/ASSETS/GRAPHIC/315_954_F4.JPEG.

[9] D. Wang *et al.*, "Ferroelectric C-Axis Textured Aluminum Scandium Nitride Thin Films of 100 nm Thickness," *IFCS-ISAF 2020 - Jt. Conf. IEEE Int. Freq. Control Symp. IEEE Int. Symp. Appl. Ferroelectr. Proc.*, pp. 2–5, 2020, doi: 10.1109/IFCS-ISAF41089.2020.9234910.

[10] U. Schroeder, M. H. Park, T. Mikolajick, and C. S. Hwang, "The fundamentals and applications of ferroelectric HfO2," *Nat. Rev. Mater.*, vol. 7, no. 8, pp. 653–669, Aug. 2022, doi: 10.1038/S41578-022-00431-2.

[11] M. H. Park *et al.*, "Ferroelectricity and Antiferroelectricity of Doped Thin HfO2-Based Films," *Adv. Mater.*, vol. 27, no. 11, pp. 1811–1831, Mar. 2015, doi: 10.1002/ADMA.201404531.

[12] Z. Liu, X. Wang, X. Ma, Y. Yang, and D. Wu, "Doping effects on the ferroelectric properties of wurtzite nitrides," *Appl. Phys. Lett.*, vol. 122, no. 12, 2023, doi: 10.1063/5.0145818.

[13]   B. Deng, Y. Zhang, and Y. Shi, "Examining the ferroelectric characteristics of aluminum nitride-based thin films," *J. Am. Ceram. Soc.*, vol. 107, no. 3, pp. 1571–1581, Mar. 2024, doi: 10.1111/JACE.19540.

[14]   F. Tasnádi *et al.*, "Origin of the anomalous piezoelectric response in wurtzite sc$_x$Al$_{1-x}$N alloys," *Phys. Rev. Lett.*, vol. 104, no. 13, pp. 1–4, Mar. 2010, doi: 10.1103/PhysRevLett.104.137601.

[15]   K. Yazawa, J. S. Mangum, P. Gorai, G. L. Brennecka, and A. Zakutayev, "Local chemical origin of ferroelectric behavior in wurtzite nitrides," *J. Mater. Chem. C*, vol. 10, no. 46, pp. 17557–17566, Dec. 2022, doi: 10.1039/D2TC02682A.

[16]   C.-W. Lee, N. U. Din, K. Yazawa, G. L. Brennecka, A. Zakutayev, and P. Gorai, "Emerging materials and design principles for wurtzite-type ferroelectrics," *Matter*, vol. 7, pp. 1644–1659, 2024, doi: 10.1016/j.matt.2024.02.001.

[17]   S. K. Ryoo *et al.*, "Fabrication of Ultrathin Ferroelectric Al$_{0.7}$Sc$_{0.3}$N Films under Complementary-Metal-Oxide-Semiconductor Compatible Conditions by using HfN$_{0.4}$ Electrode," *Adv. Mater.*, p. 2413295, 2024, doi: 10.1002/ADMA.202413295.

[18]   K.-H. Kim *et al.*, "Scalable CMOS back-end-of-line-compatible AlScN/two-dimensional channel ferroelectric field-effect transistors," *Nat. Nanotechnol.*, vol. 18, no. 9, pp. 1044–1050, Sep. 2023, doi: 10.1038/s41565-023-01399-y.

[19]   S. Fichtner, N. Wolff, F. Lofink, L. Kienle, and B. Wagner, "AlScN: A III-V semiconductor based ferroelectric," *J. Appl. Phys.*, vol. 125, no. 11, p. 114103, Mar. 2019, doi: 10.1063/1.5084945.

[20]   N. Wolff *et al.*, "Atomic scale confirmation of ferroelectric polarization inversion in wurtzite-type AlScN," *J. Appl. Phys.*, vol. 129, no. 3, 2021, doi: 10.1063/5.0033205.

[21]   K. Yazawa, C. Evans, E. Dickey, B. Tellekamp, G. L. Brennecka, and A. Zakutayev, "Low Leakage Ferroelectric Heteroepitaxial Al$_{0.7}$Sc$_{0.3}$N Films on GaN," 2024.

[22]   G. Schönweger *et al.*, "In-Grain Ferroelectric Switching in Sub-5 nm Thin Al$_{0.74}$Sc$_{0.26}$N Films at 1 V," *Adv. Sci.*, vol. 10, no. 25, p. 2302296, Sep. 2023, doi: 10.1002/ADVS.202302296.

[23]   S. Yasuoka *et al.*, "Effects of deposition conditions on the ferroelectric properties of (Al$_{1-x}$Sc$_x$)N thin films," *J. Appl. Phys.*, vol. 128, no. 11, p. 114103, Sep. 2020, doi: 10.1063/5.0015281/1077592.

[24]   S. K. Ryoo *et al.*, "Investigation of Optimum Deposition Conditions of Radio Frequency Reactive Magnetron Sputtering of Al$_{0.7}$Sc$_{0.3}$N Film with Thickness down to 20 nm," *Adv. Electron. Mater.*, vol. 8, no. 11, pp. 1–14, 2022, doi: 10.1002/aelm.202200726.

[25]   G. Schönweger *et al.*, "Ultrathin Al$_{1-x}$Sc$_x$N for Low-Voltage-Driven Ferroelectric-Based Devices," *Phys. Status Solidi - Rapid Res. Lett.*, vol. 17, no. 1, 2023, doi: 10.1002/pssr.202200312.

[26]   S. L. Tsai *et al.*, "On the thickness scaling of ferroelectricity in Al$_{0.78}$Sc$_{0.22}$N films," *Jpn. J. Appl. Phys.*,

vol. 60, no. SB, pp. 0–5, 2021, doi: 10.35848/1347-4065/abef15.

[27] R. Xu *et al.*, "Reducing Coercive-Field Scaling in Ferroelectric Thin Films via Orientation Control," *ACS Nano*, vol. 12, no. 5, pp. 4736–4743, 2018, doi: 10.1021/acsnano.8b01399.

[28] D. Wang *et al.*, "An Epitaxial Ferroelectric ScAlN/GaN Heterostructure Memory," *Adv. Electron. Mater.*, vol. 8, no. 9, pp. 1–8, 2022, doi: 10.1002/aelm.202200005.

[29] P. Michele, Z. Xuanyi, H. Bernerd, S. Pietro, and R. Matteo, "Effect of Substrate-RF on Sub-200 nm Al0.7Sc0.3N Thin Films," *Micromachines*, vol. 22, pp. 6–7, 2019.

[30] P. Wang, D. Wang, N. M. Vu, T. Chiang, J. T. Heron, and Z. Mi, "Fully epitaxial ferroelectric ScAlN grown by molecular beam epitaxy," *Appl. Phys. Lett.*, vol. 118, no. 22, p. 223504, May 2021, doi: 10.1063/5.0054539/150805.

[31] L. Chen *et al.*, "Bipolar and Unipolar Cycling Behavior in Ferroelectric Scandium-doped Aluminum Nitride," *2022 IEEE Int. Symp. Appl. Ferroelectr. Piezoresponse Force Microsc. Eur. Conf. Appl. Polar Dielectr. ISAF-PFM-ECAPD 2022*, 2022, doi: 10.1109/ISAF51494.2022.9870042.

[32] N. Wolff *et al.*, " Demonstration and STEM Analysis of Ferroelectric Switching in MOCVD-Grown Single Crystalline Al 0.85 Sc 0.15 N ," *Adv. Phys. Res.*, vol. 3, no. 5, pp. 1–8, 2024, doi: 10.1002/apxr.202300113.

[33] M. Baeumler *et al.*, "Optical constants and band gap of wurtzite $Al_{1-x}Sc_xN/Al_2O_3$ prepared by magnetron sputter epitaxy for scandium concentrations up to x = 0.41," *J. Appl. Phys.*, vol. 126, no. 4, p. 045715, Jul. 2019, doi: 10.1063/1.5101043.

[34] C. W. Lee, K. Yazawa, A. Zakutayev, G. L. Brennecka, and P. Gorai, "Switching it up: New mechanisms revealed in wurtzite-type ferroelectrics," *Sci. Adv.*, vol. 10, no. 20, p. 848, May 2024, doi: 10.1126/SCIADV.ADL0848/SUPPL_FILE/SCIADV.ADL0848_SM.PDF.

[35] N. Wolff *et al.*, "$Al_{1-x}Sc_xN$ Thin Films at High Temperatures: Sc-Dependent Instability and Anomalous Thermal Expansion," *Micromachines*, vol. 13, no. 8, p. 1282, Aug. 2022, doi: 10.3390/mi13081282.

[36] P. Wang, D. Wang, S. Yang, and Z. Mi, "Ferroelectric nitride semiconductors: Molecular beam epitaxy, properties, and emerging device applications.," in *Semiconductors and Semimetals*, vol. 114, 2023, pp. 21–69.

[37] M. L. Green *et al.*, "Fulfilling the promise of the materials genome initiative with high-throughput experimental methodologies," *Appl. Phys. Rev.*, vol. 4, no. 1, p. 011105, Mar. 2017, doi: 10.1063/1.4977487.

[38] K. R. Talley *et al.*, "Implications of heterostructural alloying for enhanced piezoelectric performance of (Al,Sc)N," *Phys. Rev. Mater.*, vol. 2, no. 6, p. 063802, Jun. 2018, doi: 10.1103/PhysRevMaterials.2.063802.

[39] J. Patidar *et al.*, "Improving the crystallinity and texture of oblique-angle-deposited AlN thin films

using reactive synchronized HiPIMS," *Surf. Coatings Technol.*, vol. 468, p. 129719, Sep. 2023, doi: 10.1016/j.surfcoat.2023.129719.

[40] B. Bakhit *et al.*, "Strategy for simultaneously increasing both hardness and toughness in ZrB2-rich Zr1−xTaxBy thin films," *J. Vac. Sci. Technol. A Vacuum, Surfaces, Film.*, vol. 37, no. 3, 2019, doi: 10.1116/1.5093170.

[41] A. Anders, "Tutorial: Reactive high power impulse magnetron sputtering (R-HiPIMS)," *J. Appl. Phys.*, vol. 121, no. 17, p. 171101, May 2017, doi: 10.1063/1.4978350.

[42] G. Greczynski, I. Petrov, J. E. Greene, and L. Hultman, "Paradigm shift in thin-film growth by magnetron sputtering: From gas-ion to metal-ion irradiation of the growing film," *J. Vac. Sci. Technol. A*, vol. 37, no. 6, p. 060801, Dec. 2019, doi: 10.1116/1.5121226.

[43] J. Patidar, K. Thorwarth, T. Schmitz-Kempen, R. Kessels, and S. Siol, "Deposition of highly-crystalline AlScN thin films using synchronized HiPIMS -- from combinatorial screening to piezoelectric devices," *Am. Phys. Soc.*, vol. 095001, pp. 1–13, 2024, doi: 10.1103/PhysRevMaterials.8.095001.

[44] G. Schönweger *et al.*, "From Fully Strained to Relaxed: Epitaxial Ferroelectric Al1-xScxN for III-N Technology," *Adv. Funct. Mater.*, vol. 32, no. 21, p. 2109632, May 2022, doi: 10.1002/ADFM.202109632.

[45] S. L. Tsai *et al.*, "Room-temperature deposition of a poling-free ferroelectric AlScN film by reactive sputtering," *Appl. Phys. Lett.*, vol. 118, no. 8, pp. 0–5, 2021, doi: 10.1063/5.0035335.

[46] R. Shibukawa, S. L. Tsai, T. Hoshii, H. Wakabayashi, K. Tsutsui, and K. Kakushima, "Influence of sputtering power on the switching and reliability of ferroelectric Al0.7Sc0.3N films," *Jpn. J. Appl. Phys.*, vol. 61, no. SH, pp. 0–5, 2022, doi: 10.35848/1347-4065/ac5db0.

[47] M. Pirro *et al.*, "Ferroelectric Considerations on Co-Sputtered 30% ALSCN with Different DC+RF Ratios," *2021 Jt. Conf. Eur. Freq. Time Forum IEEE Int. Freq. Control Symp. EFTF/IFCS 2021 - Proc.*, pp. 1–3, 2021, doi: 10.1109/EFTF/IFCS52194.2021.9604316.

[48] V. Gund *et al.*, "Towards Realizing the Low-Coercive Field Operation of Sputtered Ferroelectric ScxAl1-xN," *21st Int. Conf. Solid-State Sensors, Actuators Microsystems, TRANSDUCERS 2021*, pp. 1064–1067, 2021, doi: 10.1109/Transducers50396.2021.9495515.

[49] S. Yasuoka *et al.*, "Impact of Deposition Temperature on Crystal Structure and Ferroelectric Properties of (Al1−xScx)N Films Prepared by Sputtering Method," *Phys. Status Solidi Appl. Mater. Sci.*, vol. 218, no. 17, pp. 1–6, 2021, doi: 10.1002/pssa.202100302.

[50] T. Tominaga, S. Takayanagi, and T. Yanagitani, "Negative-ion bombardment increases during low-pressure sputtering deposition and their effects on the crystallinities and piezoelectric properties of scandium aluminum nitride films," *J. Phys. D. Appl. Phys.*, vol. 55, no. 10, 2022, doi: 10.1088/1361-

6463/ac3d5c.

[51] A. G. Chernikova *et al.*, "Improved Ferroelectric Switching Endurance of La-Doped Hf0.5Zr0.5O2 Thin Films," *ACS Appl. Mater. Interfaces*, vol. 10, no. 3, pp. 2701–2708, 2018, doi: 10.1021/acsami.7b15110.

[52] E. Yurchuk *et al.*, "Charge-Trapping Phenomena in HfO2-Based FeFET-Type Nonvolatile Memories," *IEEE Trans. Electron Devices*, vol. 63, no. 9, pp. 3501–3507, 2016, doi: 10.1109/TED.2016.2588439.

[53] M. Pešić *et al.*, "Physical Mechanisms behind the Field-Cycling Behavior of HfO2-Based Ferroelectric Capacitors," *Adv. Funct. Mater.*, vol. 26, no. 25, pp. 4601–4612, 2016, doi: 10.1002/adfm.201600590.

[54] D. K. Pradhan *et al.*, "A scalable ferroelectric non-volatile memory operating at 600 °C," *Nat. Electron. 2024 75*, vol. 7, no. 5, pp. 348–355, Apr. 2024, doi: 10.1038/s41928-024-01148-6.

[55] O. V. Pshyk, J. Patidar, M. Alinezhadfar, S. Zhuk, and S. Siol, "Beyond Structural Stabilization of Highly-Textured AlN Thin Films: The Role of Chemical Effects," *Adv. Mater. Interfaces*, p. 2400235, 2024, doi: 10.1002/ADMI.202400235.

[56] J. Patidar, O. Pshyk, L. Sommerhäuser, and S. Siol, "Low Temperature Deposition of Functional Thin Films on Insulating Substrates: Selective Ion Acceleration using Synchronized Floating Potential HiPIMS," Aug. 2024, Accessed: Sep. 03, 2024. [Online]. Available: https://arxiv.org/abs/2408.12174v1.

# Supplementary information for

# Ferroelectric AlScN thin films with enhanced polarization and low leakage enabled by high-power impulse magnetron sputtering

Federica Messi[1], Jyotish Patidar[1], Nathan Rodkey[1], Christoph W. Dräyer[2], Morgan Trassin[2], Sebastian Siol[1,*]

[1]Laboratory for Surface Science and Coating Technologies,
Empa – Swiss Federal Laboratories for Materials Science and Technology, Switzerland

[2]Department of Materials, ETH Zürich, Switzerland

E-mail: sebastian.siol@empa.ch

**SI 1**: hysteresis study

The $Al_{1-x}Sc_xN$ depositions are conducted in reactive sputtering mode by introducing nitrogen as a reactive gas along with argon as the inert sputtering gas. During $Al_{1-x}Sc_xN$ deposition, both argon (Ar) and nitrogen ($N_2$) gases are introduced into the chamber. The hysteresis loops in **Figure S1** are measured by varying the ratio between the Ar flow and the $N_2$ flow maintaining the total flux of 32 sccm and a constant pressure. This is done for both Al and Sc to identify when the compound mode begins. The conditions selected for the subsequent experiments are 12 sccm of $N_2$ and 20 sccm of Ar, achieving complete target poisoning for both targets.

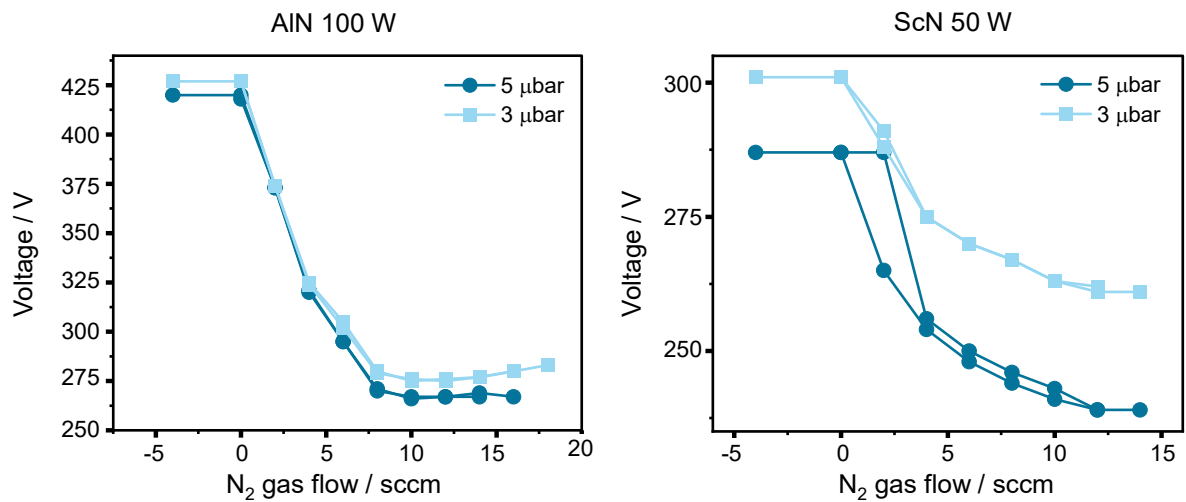

*Figure S1. Measurement of hysteresis for Al and Sc using DCMS.*

**SI 2:** set up of the chamber for the deposition and schematic of samples distribution and workflow

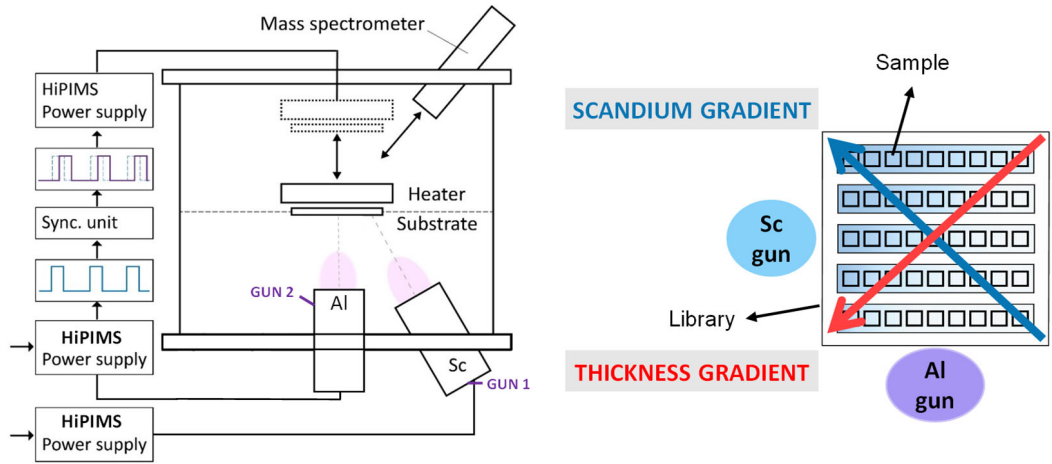

***Figure S2.*** *Illustration of the chamber for combinatorial libraries deposition; the Al and Sc target are positioned next to each other while the substrate is kept fixed during the deposition. This results in both thickness and scandium gradients on the library. On each library a 9 x 5 grid defines 45 samples with different values of scandium concentration and thickness.*

**SI 3:** characterization results

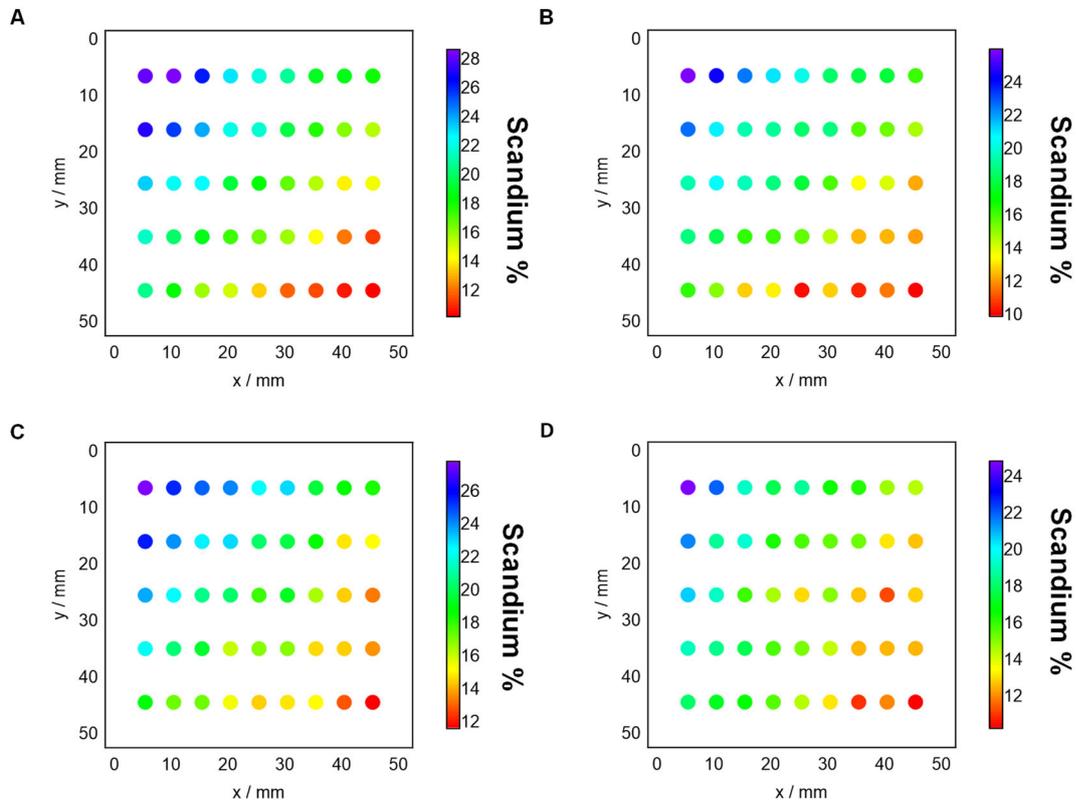

***Figure S3.*** *XPS results of Scandium concentration for the libraries:* ***a.*** *grounded;* ***b.*** *-5 V substrate bias;* ***c.*** *-10 V substrate bias;* ***d.*** *-20 V substrate bias. The composition ranges from 10% to 28% scandium.*

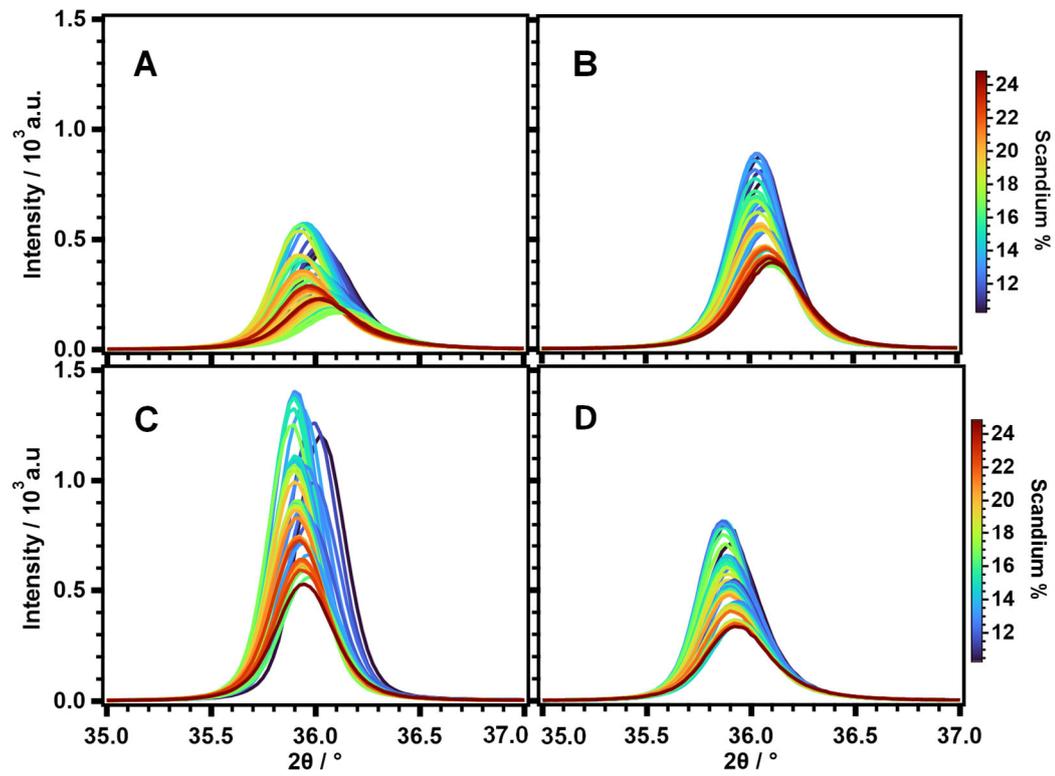

*Figure S4.* XRD θ-2θ measurement of the Al$_{1-x}$Sc$_x$N (002) peak as a function of scandium to evaluate the crystal orientation of the films: **a.** grounded; **b.** -5 V substrate bias; **c.** -10 V substrate bias; **d.** -20 V substrate bias. The highest intensity is reported for the library deposited with -10 V substrate bias.

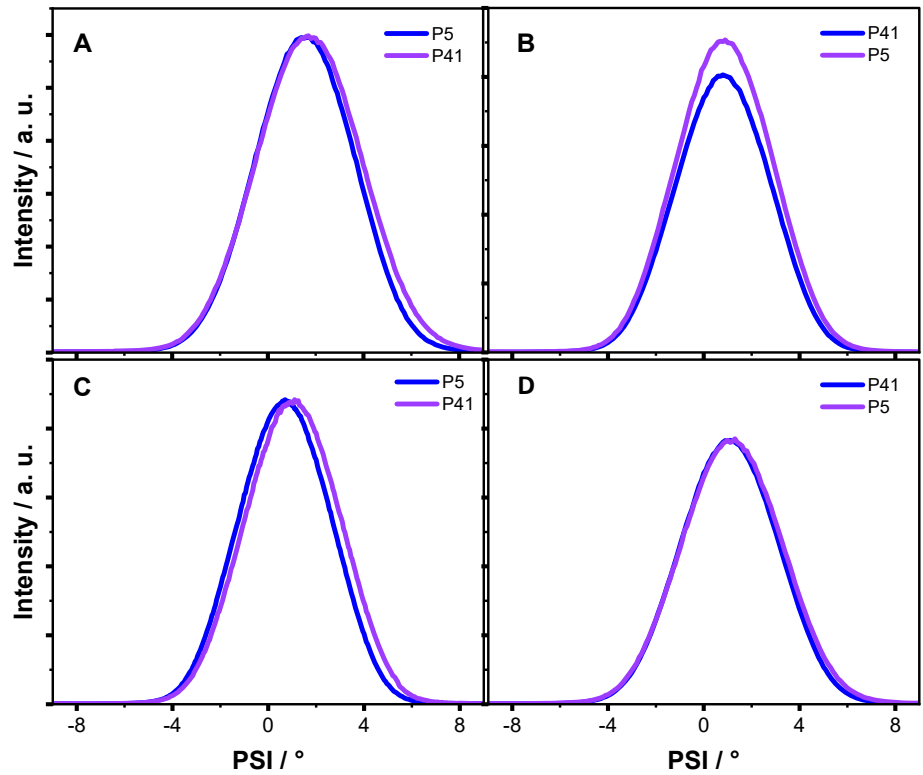

*Figure S5.* XRD psi scans on the Al$_{1-x}$Sc$_x$N (002) peak for two samples on the extreme edges of combinatorial library (point 5 and point 41) to evaluate the misorientation of the grains: **a.** grounded; **b.** -5 V substrate bias; **c.** -10 V substrate bias; **d.** -20 V substrate bias. The peak occurs always around 1°, confirming the growth of the grains along the c-axis.

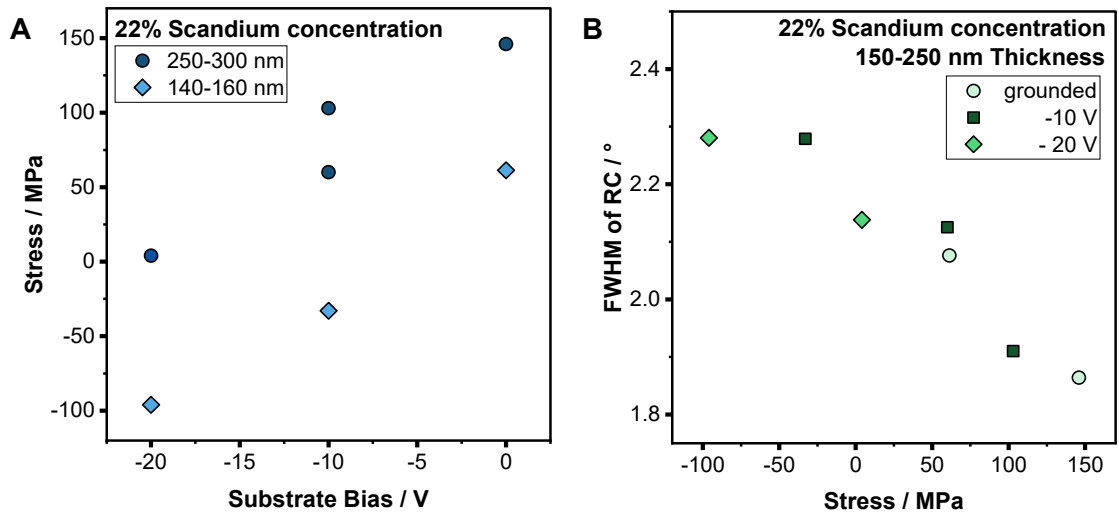

***Figure S6.*** *Stress analysis on silicon: **a.** dependence on the substrate bias and film thickness; **b.** dependence on the FWHM of the rocking curve of the $Al_{1-x}Sc_xN$ peak. In the films deposited on Si, the stress is mainly influenced by factors such as film thickness, substrate bias, and FWHM. Thicker films exhibit tensile stress, while as the FWHM increases, the stress transitions from tensile to compressive. When examining the impact of substrate bias, the trend is as expected: increasing the negative bias leads to greater ion bombardment, which results in more compressive stress in the film*

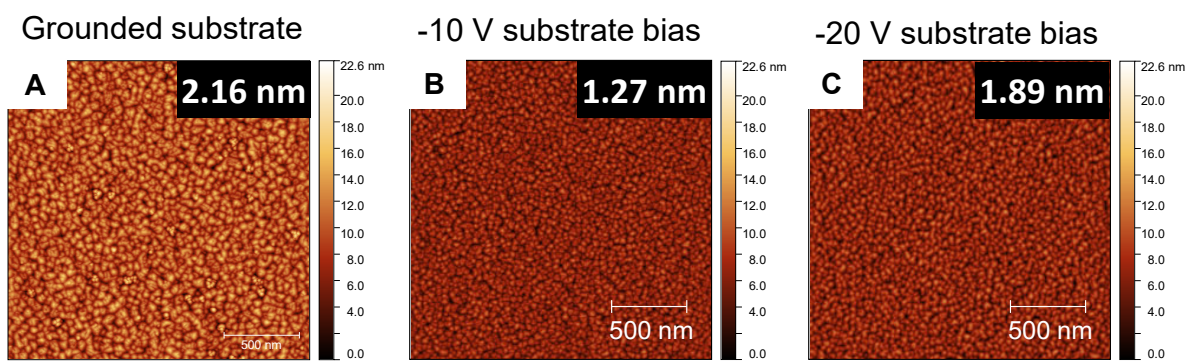

***Figure S7.*** *AFM imaging of 250 nm $Al_{0.78}Sc_{0.22}N$ films deposited on Si with grounded or sync-bias, and substrate rotation. All films exhibit limited RMS roughness, with the sample deposited at -10 V demonstrating the best surface quality.*

*Table S1.* Results of the ferroelectric switching of the samples belonging to the grounded library and the library with -10 V substrate bias. The leakage current refers to the average between the positive and negative values calculated at the coercive field.

| Bias / V | Sc % | $E_c$ / MV cm$^{-1}$ | $P_r$ / µC cm$^{-2}$ | Leakage current density / A cm$^{-2}$ |
|---|---|---|---|---|
| Grounded | 14.3 | 5.5 | 154.6 | 0.19 |
| Grounded | 15.9 | 5.4 | 157.9 | 0.15 |
| Grounded | 16.6 | 5.4 | 169.4 | 0.22 |
| Grounded | 16.8 | 5.1 | 149.1 | 0.18 |
| Grounded | 17.4 | 5.1 | 163.4 | 0.11 |
| Grounded | 18.5 | 5.1 | 156.8 | 0.18 |
| Grounded | 20.2 | 5.1 | 142.6 | 0.11 |
| Grounded | 21.6 | 5.0 | 172.0 | 0.19 |
| Grounded | 22.2 | 5.0 | 145.0 | 0.12 |
| Grounded | 22.6 | 5.0 | 153.9 | 0.12 |
| -10 V | 16.5 | 5.9 | 162.3 | 0.20 |
| -10 V | 17 | 5.9 | 170.7 | 0.22 |
| -10 V | 19.4 | 5.7 | 158.8 | 0.2 |
| -10 V | 19.5 | 5.5 | 155.2 | 0.21 |
| -10 V | 20.2 | 5.3 | 165.7 | 0.22 |
| -10 V | 22.4 | 5.0 | 156.7 | 0.18 |

**SI 4**: Piezoresponse Force Microscopy

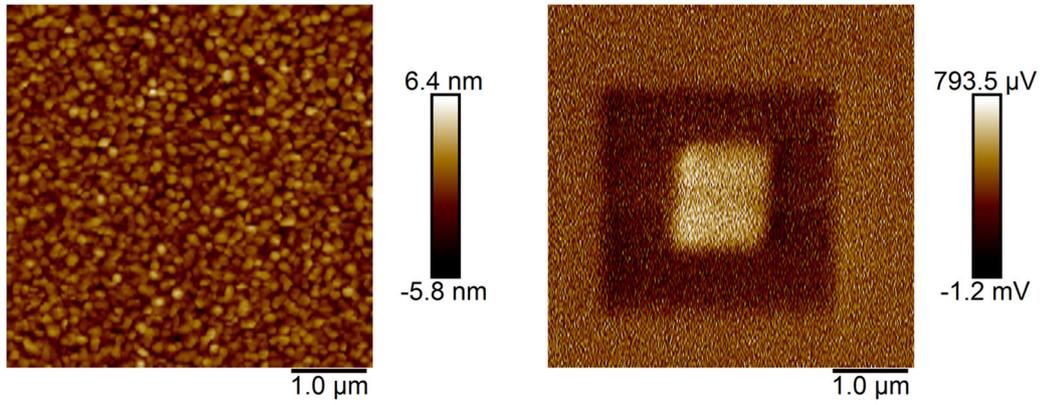

*Figure S8.* PFM results of the 10 nm thick film, showing an apparent change in the polarization state.

**SI 5:** deposition methods of the single composition samples and characterization results

Isolated samples with homogenous scandium concentrations are deposited in order to evaluate the effect of stress on the properties of the samples. For the deposition of these isolated samples, the configuration of the targets is modified. The Al and Sc guns are positioned opposite to one another and a closed magnetic configuration is used. The substrate is rotated at 40 rpm during the entire deposition to achieve homogeneous composition and thickness. The single-composition samples are deposited on both 13x13 mm$^2$ silicon. Nitrogen was routed directly to the targets to ensure easier poisoning of the targets while the Ar gas was supplied away from the target in the chamber. The flow rates of Ar/N$_2$ and working pressure are maintained constant at 20 and 12 sccm, respectively. The substrate was either grounded or a synchronized bias pulse of −10 V and -20 V was applied. The target scandium composition was set at 22%, with two thicknesses chosen: 250 nm and 150 nm. The layered structure was consistent with the library layered deposition, with the only difference being the platinum layer deposition, which was performed using HiPIMS instead of DCMS, and the Pt top contacts, which were deposited with diameters of 50 µm, 100 µm, 200 µm and 400 µm.